# Second harmonic generation in atomically thin MoTe$_2$

*Yu Song, Ruijuan Tian, Jiulong Yang, Rui Yin, Jianlin Zhao[\*], Xuetao Gan[\*]*

*MOE Key Laboratory of Material Physics and Chemistry under Extraordinary Conditions, and Shaanxi Key Laboratory of Optical Information Technology, School of Science, Northwestern Polytechnical University, Xi'an 710072, China*

[\*]Corresponding author: xuetaogan@nwpu.edu.cn; jlzhao@nwpu.edu.cn

**Abstract**

We have studied on optical second harmonic generations (SHGs) from atomically thin MoTe$_2$ flakes with 2H and 1T′ phases. From 2H-MoTe$_2$ samples with odd (even) numbers of layers, strong (negligible) SHGs are observed due to the layer-dependent broken inversion symmetry. When pumped by a telecom-band laser, SHG from a monolayer 2H-MoTe$_2$ is about one order of magnitude stronger than that from a monolayer WS$_2$; an extremely high second-order nonlinear susceptibility of 2.5 nm/V is estimated, presenting the highest value among those reported in two-dimensional materials. SHG measurements in MoTe$_2$ are also demonstrated as an efficient way to distinguish the 2H-to-1T′ phase transition. Comparing to the SHG in 2H-MoTe$_2$, 1T′-MoTe$_2$'s SHG has much lower efficiency and the polarization dependence is changed from six-fold to two-lobe pattern.

**1. Introduction**

Two-dimensional (2D) transition metal dichalcogenides (TMDs) with few atom-layers exhibit fascinating layer-number dependent electronic and optical properties.[1,2] For instance, some TMDs present indirect to direct bandgap transition when they are thinned down to monolayer, opening up a new window for 2D material-based photonics and optoelectronics in the spectral rang from near-infrared to visible.[3-5] The low dimensionality of 2D TMDs gives rise to reduced dielectric screening of Coulomb interactions between charge carriers, enabling strong light-matter interactions, ultrafast radiative recombination rate, and exotic excitonic effect.[6-9] In addition, the electronic band structures of 2D TMDs could be described by two copies of degenerated conduction and valence bands around the K and K′ points, which produces valley- and spin-dependent optical and electrical properties.[10-13] A variety of interesting optoelectronic devices have been developed based on 2D TMD semiconductors, including photodetectors with low dark current and high detectivity, and light emitting diodes with high external quantum efficiency.[14-21]

Nonlinear optical responses in 2D TMDs is another intriguing attribute for potentially extending their optoelectronic applications. Specifically, because of the broken inversion symmetry in few-layer $MX_2$ (M=Mo, W; X=S, Se) flakes with odd layer thicknesses, considerable second harmonic generations (SHGs) are observed, though their bulk materials were well recognized without second-order nonlinearity.[22-26] It is also possible to design spiral nanostructures during the material growth, which not only maintains the broken symmetry in each monolayer but also increases the effective material thickness to greatly strengthen SHGs in 2D TMDs.[27] When the two-photon energy of the pump laser is on-resonance with the exciton of monolayer $WSe_2$, a high second-order nonlinear susceptibility of 1,000 pm/V is estimated, which is about three orders of magnitude higher than those in conventional bulk materials.[28] Combining with the electrical tunability and valley selectivity of the strong excitons in monolayer $MX_2$, this exciton-enhanced SHG could be modulated by an order of magnitude with an application of a vertical electrical field, which is also verified to have counter-circular polarization to the pump laser.[29] The extraordinary and tunable SHGs in 2D TMDs provide new possibilities to construct nonlinear optoelectronic devices, including nonlinear electro-optic modulators, coherent light source generators, etc.

Most of the SHG studies of 2D TMDs were implemented on mono- and few-layer $MoS_2$, $MoSe_2$, and their tungsten analogs. In this paper, we report the measurements of SHGs in mono- and few-layer $MoTe_2$, which has recently arisen as an appealing 2D TMD semiconductor for photonic and electronic devices. In contrast to the $MoS_2$ and $MoSe_2$, the indirect-to-direct bandgap crossover in atomically thin $MoTe_2$ occurs before reaching monolayer thickness.[30] And the photoluminescence emission and excitonic absorption are located in the near-infrared range around 1.1 eV, bridging the comparatively large bandgap of other monolayer TMDs and zero-bandgap graphene. Another important property of $MoTe_2$ is its large exciton-binding energy around 0.6 eV,[31] providing a unique opportunity for achieving the first 2D material-based nanolaser at room temperature.[32] The spin-orbit coupling in $MoTe_2$ is much stronger than that in $MoS_2$ or $MoSe_2$, which could contribute to a longer decoherence time for exciton valley and spin indexes and new valleytronic devices.[33] Those intriguing attributes are revealed in the semiconducting hexagonal (2H)-$MoTe_2$. Note that stable $MoTe_2$ could also exist in a semimetal monoclinic (1T′) phase, which endows more opportunities for electronic and optoelectronic applications. 2H-to-1T′ phase changes in $MoTe_2$ could be experimentally realized through laser irradiation, electrostatic gating, and thermal synthesis, and the distinctly modulated optical and electrical properties were also demonstrated for functional devices.[34-36]

Here, we study on SHGs from few-layer $MoTe_2$ with different phases. We observe strong SHGs from 2H-$MoTe_2$ with odd numbers of layers, which is consistent with the broken inversion symmetry of the crystal structure. When pumped by a telecom-band pulsed laser, SHG from the monolayer 2H-$MoTe_2$ is almost one order of magnitude stronger than that from monolayer WS$_2$. Its high second-order nonlinear susceptibility $\chi^{(2)}$ is estimated as 2.5 nm/V, which is the highest one among those reported in other 2D materials. Zero SHG in bilayer 2H-MoTe$_2$ is verified, though the indirect-to-direct bandgap crossover is not clear between monolayer and bilayer. By laser irradiation, a MoTe$_2$ flake with 2H phase is changed into 1T′ phase, presenting a significantly decreased SHG by two orders of magnitude. SHGs' polarization dependences are exploited to identify the phase transition.

## 2. Materials and methods

The atomically thin 2H-MoTe$_2$ crystals are prepared by mechanically exfoliating a synthetic semiconducting bulk material (supplier: HQ Graphene) onto silicon substrates covered with 280 nm thick silicon dioxide (SiO$_2$/Si). The few-layer MoTe$_2$ flakes are identified using an optical microscope. As shown in Figure 1(a), flakes with different thicknesses are obtained with various optical contrasts. Atomic force microscope (AFM) technique is further employed for confirming the layer numbers. Figure 1(b) displays the AFM image of the MoTe$_2$ region marked in the white box of Figure 1(a). The inset indicates the layer heights along the dashed line, covering mono-, bi- and tri-layers with the corresponding thicknesses of 0.8, 1.5, and 2.6 nm, respectively.

SHG measurements of the few-layer MoTe$_2$ are implemented in a home-built vertical microscope setup with the reflection geometry. A fiber-based pulsed laser is chosen as the fundamental pump radiation, which has a central wavelength around 1550 nm, a repetition rate of 18.5 MHz, and a pulse width of 8.8 ps. A 50× microscope objective lens with a numerical aperture of 0.75 is employed to focus the pump laser into a spot size of about 2 μm on the sample. The second harmonic (SH) signal scattered from the MoTe$_2$ sample is collected by the same objective lens. In the signal collection path, a dichroic mirror is used to filter out the pump laser from the SH signal, which is finally analyzed and detected by a spectrometer mounted with a cooled silicon charge-coupled device (CCD) camera. To study the polarization dependence of the SH radiation, we realize linearly polarized pump lasers along different polarization directions by passing a circularly polarized pump laser through a rotated polarizer. Another polarizer is placed in the signal collection path, whose direction is rotated correspondingly to the pump polarization to collect the parallel or perpendicular components of the SH signal. For SHG spatial mapping, samples are placed on a piezo-actuated stage for the in-plane scanning, while the excitation and collection light-spots are fixed. Simultaneously, the optical power integrated over the whole SH spectral linewidth is recorded by a sensitive photomultiplier tube to reveal SHG's position dependence.

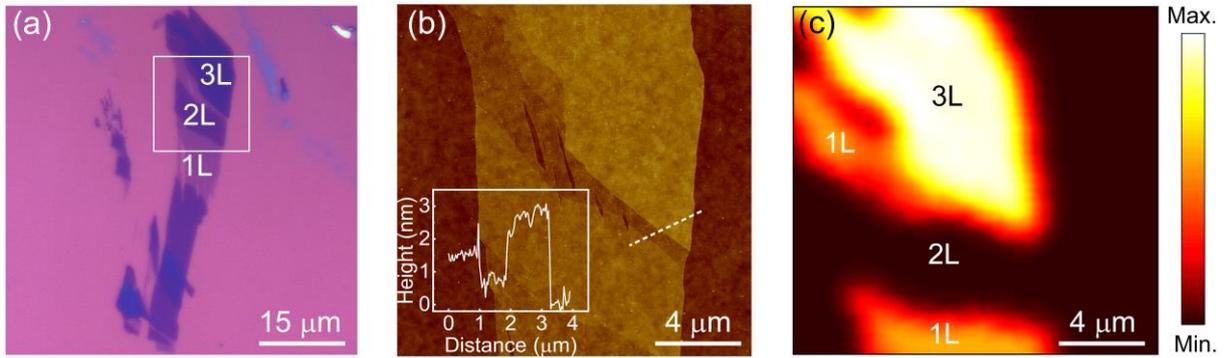

**Figure 1**. (a) Optical microscope image of a few-layer $MoTe_2$ sample exfoliated on a 280 nm $SiO_2$/Si. (b) AFM image of the flakes marked in the white box in (a). The inset shows the thicknesses measured along the white dashed line, indicating the mono-, bi- and tri-layers with corresponding thicknesses of 0.8, 1.5, and 2.6 nm, respectively. (c) SHG spatial mapping image of the $MoTe_2$ flakes displayed in (b).

## 3. Results and discussions

3.1 Layer-dependent SHGs in few-layer $MoTe_2$

We first measure SHGs from the prepared sample shown in Figure 1(a) with an averaged pump power lower than 1 mW to avoid the laser-induced phase transition.[34] A considerable SH signal around the wavelength of 775 nm is observed when the pump laser is focused on the monolayer area. By changing the pump power gradually, the SHG's power dependence is characterized, showing an expected quadratic function governed by the process that two photons of the pump laser are converted into one photon in SH signal. However, no any SH signal is detected in the bilayer MoTe$_2$ even when the pump power is increased to 10 mW, which is the maximum power without any observations of phase transitions. When the laser is focused on the trilayer region, a SH signal stronger than that from monolayer is obtained. A more straightforward result indicating SH signals of different layers is shown in Figure 1(c), which is a SHG spatial mapping over the region shown in Figure 1(b). Well-defined regions with varied SH intensities are obtained. Strong SH signals are observed over the mono- and tri-layer regions, and the bi-layer flakes show contrastingly zero SHG.

To further study layer-dependent SHGs from few-layer MoTe$_2$, more MoTe$_2$ samples are prepared and measured. The optical microscope image of one of the samples is displayed in Figure 2(a). It has multiple thick layers, and the layer numbers are confirmed by AFM. A SHG spatial mapping is implemented over the region marked in the white box in Figure 2(a), as shown in Figure 2(b). For these thicker few-layer MoTe$_2$, strong SH signals occur in odd numbers of layers, while even numbers of layers do not show detectable SHG. We also measure that SH signal from the bulk material is negligible. Strong SHGs in odd-layer 2H-MoTe$_2$ indicate the broken inversion symmetry in their crystal structures, which is same as those demonstrated in few-layer 2H-MoS$_2$. We plot the relative SH intensities for different layer thicknesses in Figure 2(c). For each measurement of SH intensity, the polarization of the pump laser is optimized with respect to the lattice orientation of MoTe$_2$ to achieve the maximum SH signal. We attribute the SHGs' layer-dependences to the absorption of SH signal by the few-layer MoTe$_2$ flakes. The photon energy of the SH signal is about 1.6 eV, which is much larger than the optical bandgap of few-layer MoTe$_2$. We neglect the optical interference effect between the material interfaces considering the ultrathin film for the layer number smaller than ten. As reported in Ref. [30], MoTe$_2$ has direct bandgap for the mono- and bi-layers, and transits to the indirect bandgap for the tri- and tetra-layers. Hence, the optical absorption of SH signal in monolayer would be stronger than that in trilayer,[37] resulting in higher collection efficiency of SHG in trilayer MoTe$_2$. After the layer number increases to be more than five, the indirect bandgap weakens the light absorption, which therefore induces a much stronger SHG in the five-layer flake. For even thicker flakes, while the optical absorption in each layer is still weak for the indirect bandgap, the total attenuation over the SH signal is strong caused by the more layer numbers. The SH intensities decrease for the flakes with layer numbers more than five.

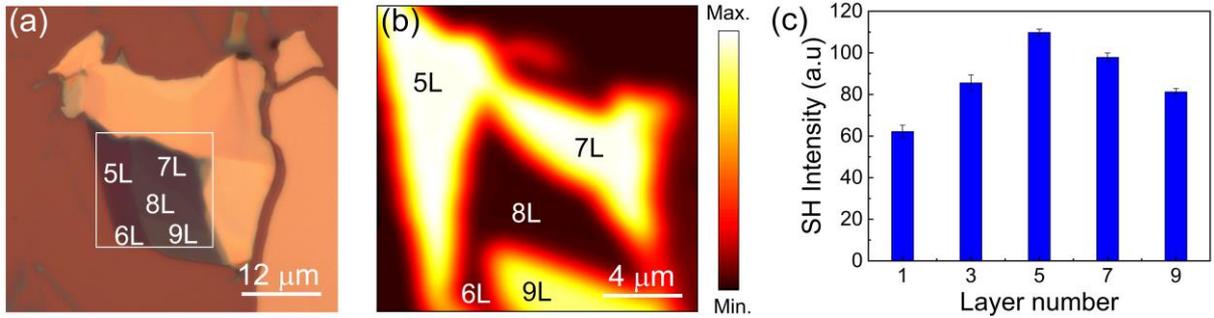

**Figure 2**. (a) Optical microscope image of a thicker MoTe$_2$ sample exfoliated on a 280 nm SiO$_2$/Si. (b) SHG spatial mapping image of the MoTe$_2$ flake marked by the white box in (a). (c) Layer dependences of MoTe$_2$'s SH intensities, where the uncertainties are indicated by the error bars.

3.2 Polarization-dependent SHGs in monolayer MoTe$_2$

The polarization dependences of SHGs from the few-layer MoTe$_2$ are studied further, which could indicate the crystalline symmetry and sample orientation. Figure 3(a) shows one of the acquired results from a monolayer MoTe$_2$, where the red dots (black squares) represent the parallel (perpendicular) component of SH radiations. For both of the two components, the polarization-resolved SH intensities of 2H-MoTe$_2$ exhibit strongly varied, six-fold symmetric responses as a function of the azimuthal angle rotating about its surface normal. It directly reveals the underlying symmetry and orientation of monolayer MoTe$_2$. To illuminate that, we schematically plot the structure of monolayer MoTe$_2$ in Figure 3(b). It contains a sheet of Mo atoms with a three-fold coordinate symmetry, which is sandwiched between two hexagonal planes of Te atoms. These Te-Mo-Te units, which we refer to as monolayer, are connected together by weak van der Waals forces and are stacked with a 2H-type symmetry. Monolayer

MoTe$_2$ is expected to belong to the $D_{3h}^1$ space group. For this case, the second-order nonlinear susceptibility tensor has a single nonzero element: $\chi^{(2)} \equiv \chi^{(2)}_{yyy} = -\chi^{(2)}_{xxy} = -\chi^{(2)}_{yxx} = -\chi^{(2)}_{xyx}$, where $x$, $y$, and $z$ are the crystalline coordinates. Here, $x$ axis is defined along the armchair direction of the hexagonal plane, which is 30° from the zigzag direction. In our experiments, the incident beam is excited along the $z$ direction. We define an angle $\theta$ between the $x$ direction and the polarization of the fundamental pump laser. It is straightforward to obtain that the parallel and perpendicular components of the SH field are proportional to $\sin 3\theta$ and $\cos 3\theta$, respectively. The angle dependences of SH intensity can be described as

$$\begin{aligned} I_\parallel &= I_0 \cos^2(3\theta) \\ I_\perp &= I_0 \sin^2(3\theta) \end{aligned} \tag{1}$$

Here, $I_0$ is the maximum intensity of the SH response. We fit the SHG's polarization-dependent data with the above functions, which agree well with the expectations.

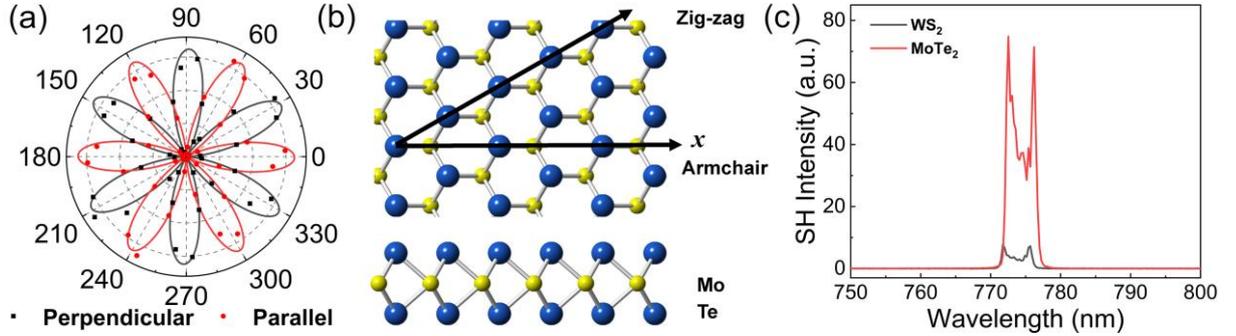

**Figure 3**. (a) Polar plot of the SH intensities from monolayer MoTe$_2$. The SH radiation components detected parallelly and perpendicularly to the polarization of the pump laser are shown. (b) Top and side view of monolayer MoTe$_2$'s lattice structure. (c) SH spectra from the monolayer MoTe$_2$ and the monolayer WS$_2$ when pumped by the same laser at the wavelength around 1550 nm.

3.3 Second-order nonlinear susceptibility of monolayer MoTe$_2$

To estimate the absolute value of $\chi^{(2)}$ in monolayer MoTe$_2$, we compare its SH intensity to that from the surface of a $z$-cut bulk crystal of lithium niobate (LN), which can be expressed as:[38]

$$\chi^{(2)}_{\text{MoTe}_2} = \frac{1}{16\pi\Delta k_{\text{LN}}\Delta h} \frac{[n_{\text{LN}}(\omega)+1]^3}{n_{\text{LN}}(\omega)n^{1/2}_{\text{LN}}(2\omega)} \left(\frac{I_{\text{MoTe}_2}(2\omega)}{I_{\text{LN}}(2\omega)}\right)^{1/2} \chi^{(2)}_{\text{LN}} \qquad (2)$$

where $\Delta h \sim 0.8$ nm is the thickness of the monolayer MoTe$_2$ determined by AFM. $\chi^{(2)}_{\text{LN}}$ is LN's second-order nonlinear susceptibility, and $n_{\text{LN}}(\omega)$ and $n_{\text{LN}}(2\omega)$ denote the linear refractive indices of LN at the fundamental and SH frequencies, respectively. $\Delta k_{\text{LN}}$ is the frequency-dependent phase mismatch in LN between the fundamental laser and SH signal. By counting $\Delta k_{\text{LN}}$ and $\chi^{(2)}_{\text{LN}}$, $\chi^{(2)}_{\text{MoTe}_2}$ is estimated as 2.5 nm/V. To the best of our knowledge, this is the highest value among those reported in other 2D TMDs when pumped by a telecom-band laser. To verify this high $\chi^{(2)}$, we also measure SHG from a chemical vapor deposition (CVD) grown monolayer WS$_2$ and make a comparison with that from monolayer MoTe$_2$. In the measurement, the two samples are excited with the same pump laser at the wavelength around 1550 nm, and the laser polarizations are optimized to achieve their maximum SH intensities. With the same pump power, SH spectra from the two samples are acquired, as shown in Figure 3(c). Here, the SH spectra have no Gaussian profile since the non-Gaussian spectral lineshape of the fiber-based picosecond laser. We observe that the SH signal from the monolayer MoTe$_2$ is almost one order of magnitude stronger than that from the monolayer WS$_2$. As have been widely reported, monolayer WS$_2$ is one of 2D materials possessing strongest second-order nonlinear process, which thence confirms the high $\chi^{(2)}$ in monolayer MoTe$_2$. Because both the fundamental wave and SH signal are off-resonance from monolayer MoTe$_2$'s excitons,[39] the obtained high $\chi^{(2)}$ is not governed by the specific exciton resonance. Considering the stronger second-order nonlinearity in compounds of telluride than that in compounds of sulfide,[40,41] we expect the high $\chi^{(2)}$ in monolayer MoTe$_2$ is its intrinsic material behavior, which promises the construction of nonlinear optoelectronic devices.

3.4 Identifying MoTe$_2$'s phase transition by SHGs

In the recently reported 2H-to-1T′ phase changes in MoTe$_2$,[34-36] the verification methods of phase transitions mainly rely on Raman spectra, metal contacts, and transmission electron microscope images. SHG is an excellent tool in studying crystal structure and symmetry. The demonstrated strong and polarization-dependent SHGs in MoTe$_2$ could be considered as an efficient way to identify its phase transition. Here we carry out laser ablation on multilayer MoTe$_2$ flakes to make the phase transition, as demonstrated in Ref. [34]. A multilayer 2H-MoTe$_2$ (~10 nm) is exfoliated onto the 280 nm SiO$_2$/Si substrate and loaded onto the SHG measurement setup. By increasing the power of the picosecond pulsed laser to a value much higher than 10 mW, distinct laser ablation over the MoTe$_2$ layer could be observed from the imaging camera of the measurement setup. Figures 4(a) and (b) display the optical microscope images of the MoTe$_2$ sample before and after the laser ablation, where the ablation region is marked by the white box. To get a continuous ablation region, the sample is spatially scanned in-plane relative to the laser spot by the piezo-stage. During the laser ablation, the MoTe$_2$ layer is etched gradually layer-by-layer because of the heating. Also, the 2H-to-1T′ phase change in MoTe$_2$ will yield a considerable refractive index variation since the transition from semiconductor to semimetal. Therefore, in the optical microscope images, the ablation region has a remarkably varied optical contrast. The lattice structures of 2H- and 1T'-MoTe$_2$ are given in the insets of Figures 4(a) and (b), respectively. The topography of the laser-irradiated region is then studied with the AFM technique. The surface of the laser-irradiated region is rougher (with surface RMS changed from 1.82 to 6.55 nm) than the pristine crystal, as shown in Figure 4(c).

SHG measurements are carried out over the interesting MoTe$_2$ region before and after the laser ablation. A significant reduction of the SH intensities by near two orders of magnitude is observed after the laser ablation. The polarization dependences of the SHGs are characterized separately as well by measuring their parallel components. Because of the $D_{3h}^1$ space group in pristine 2H-MoTe$_2$, an expected six-fold pattern is obtained, as shown in Figure 4(d). After the laser ablation, the polarization-dependence of MoTe$_2$'s SHG changes significantly, presenting a two-lobe pattern in Figure 4(e). The important contrast of SHGs' polarization dependences in the MoTe$_2$ crystal before and after laser ablation could be attributed to the phase transition between them. The bulk 1T′-MoTe$_2$ is known to be distorted monoclinic, which belongs to the $C_{2h}^2$ space group and has inversion symmetry. However, if the bulk 1T′-MoTe$_2$ is reduced down to few-layer, strong SHG could be observed in the even numbers of layers due to the broken inversion symmetry.[42] And the crystal structure of the even layer 1T′-MoTe$_2$ belongs to $C_s^1$ space group, as indicated by the inset of Figure 4(b), whose second-order nonlinear susceptibility tensor governs a SHG's polarization-dependence consistent with the experimental results shown in Figure 4(e).

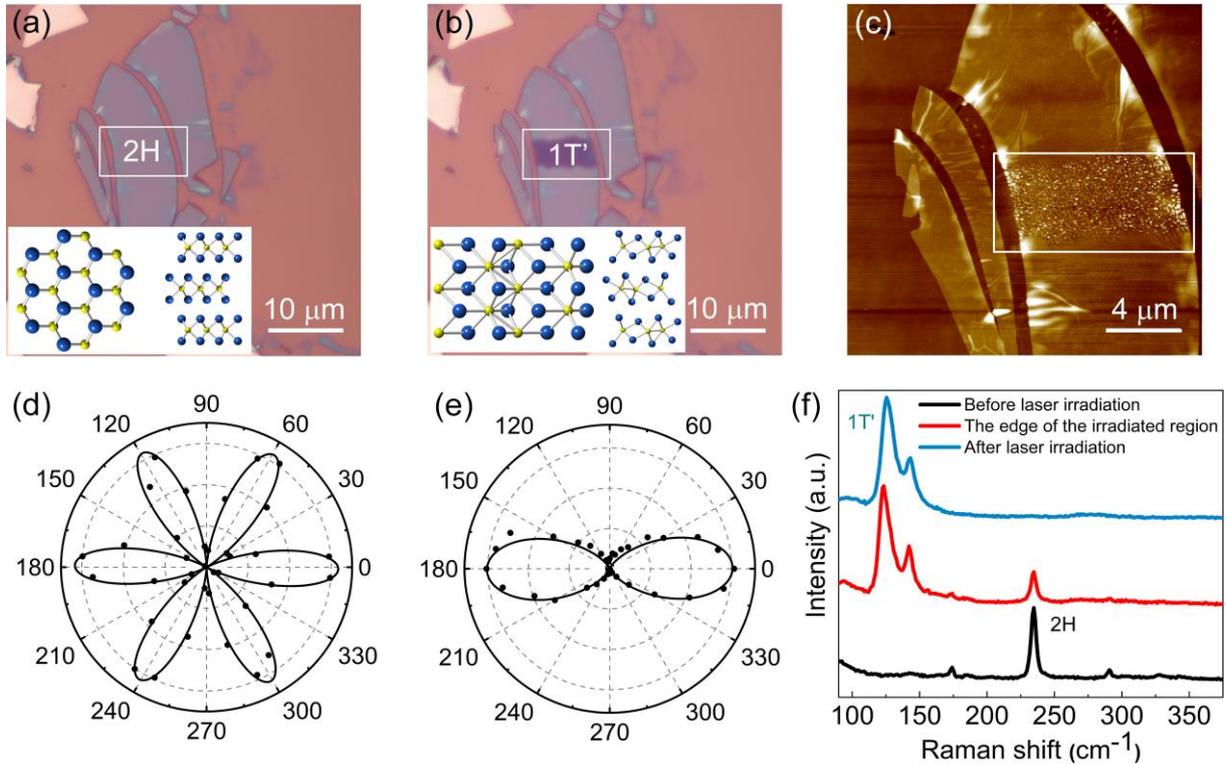

**Figure 4.** (a-b) Optical microscope images of a mechanically exfoliated 2H-MoTe$_2$ flake (a) before and (b) after laser irradiation over an area marked by the white box. Insets show the lattice structures of 2H- and 1T′-MoTe$_2$. In each inset, the left image is the top view of a monolayer, and the right image is the side view of a trilayer. (c) AFM image of the MoTe$_2$ flake after laser irradiation. (d-e) Polar plots of the SH intensities from the few-layer MoTe$_2$ (d) before and (e) after laser irradiation. (f) Raman spectra taken at the regions with and without laser irradiation as well as their interfacing region, indicating different Raman signatures of 2H- and 1T′-MoTe$_2$.

To validate the phase transition recognized by the polarization-dependent SHGs, Raman spectra are further taken from the regions with and without laser irradiation as well as their interfacing region of the sample, as shown in Figure 4(b). The measurements are implemented in a home-built Raman microscope with a 532 nm pump laser. The pump power is controlled smaller than 100 µW to avoid the further laser ablation. The Raman signal is analyzed with a spectrometer mounted with a cooled silicon CCD. The acquired results are displayed in Figure 4(f). Evidence of phase transition from 2H to 1T′ is obtained from the Raman spectra of the laser irradiation region. The initial 2H-MoTe$_2$ flake exhibits distinct Raman modes near 174, 235 and 291 cm$^{-1}$, whereas the laser-irradiated area shows new peaks near 124, 142, and 272 cm$^{-1}$, featuring a signature of 1T′-MoTe$_2$.[34] In the interfacing region, both 2H and 1T′ phases coexist. Similar SHG measurements on other several laser-irradiated 2H-MoTe$_2$ few-layers are implemented, presenting similar weakened intensity and varied polarization-dependence as well. These measurements indicate the polarization-dependent SHG could be considered as a straightforward optical method for determining phase transition in 2D materials.

## 4. Conclusions

In conclusion, we report the observations of strong (negligible) SHGs in atomically thin 2H-MoTe$_2$ with odd (even) numbers of layers, determined by the layer-dependent broken inversion symmetry. SHG in monolayer 2H-MoTe$_2$ is almost one order of magnitude stronger than that in monolayer WS$_2$, indicating its strongest second-order nonlinearity among those reported in other 2D materials when pumped by a telecom-band laser. This extra-strong SHG in few-layer 2H-MoTe$_2$ may open up new windows of their optoelectronic applications in nonlinear regime. Relying on the possibility of laser-induced 2H-to-1T′ phase transition in MoTe$_2$, its SHG is also employed to distinguish the phase transition by the polarization-dependence.

**Acknowledgements**

The National Natural Science Foundations of China (61522507, 61775183, 11634010), the Key Research and Development Program (2017YFA0303800), the Key Research and Development Program in Shaanxi Province of China (2017KJXX-12) and the Fundamental Research Funds for the Central Universities (3102017jc01001).